# Potential root mean square error skill score

Martin János Mayer[a,*], Dazhi Yang[b]

[a] Department of Energy Engineering, Faculty of Mechanical Engineering, Budapest University of Technology and Economics, Műegyetem rkp. 3, H-1111, Budapest, Hungary

[b] School of Electrical Engineering and Automation, Harbin Institute of Technology, Harbin, Heilongjiang, China

*Corresponding author. *Email address*: mayer@energia.bme.hu

**Abstract**

Consistency, in a narrow sense, denotes the alignment between the forecast-optimization strategy and the verification directive. The current recommended deterministic solar forecast verification practice is to report the skill score based on root mean square error (RMSE), which would violate the notion of consistency if the forecasts are optimized under another strategy such as minimizing the mean absolute error (MAE). This paper overcomes such difficulty by proposing a so-called "potential RMSE skill score," which depends only on: (1) the crosscorrelation between forecasts and observations, and (2) the autocorrelation of observations. While greatly simplifying the calculation, the new skill score does not discriminate inconsistent forecasts as much, e.g., even MAE-optimized forecasts can attain a high RMSE skill score.

## 1    Introduction

Proper and insightful forecast verification is a key element of most studies dealing with forecasting. The question of which metrics are the best to use for the verification is a subject of decades-long debate. That said, there is still a lack of a unified verification practice in many forecasting fields, making it difficult to evaluate and compare the goodness of novel forecasting methods presented in different papers. Murphy and Winkler [1] proposed a distribution-oriented verification framework based on the joint distribution of forecasts and observations and the different factorizations of the mean square error (MSE). The Murphy-Winkler verification framework is still in use in several forecasting disciplines; e.g., it became the main recommendation for a standardized deterministic solar forecasting verification practice by 33 prominent researchers in the field of solar energy meteorology.[2] According to Gneiting,[3] the most commonly used scoring function in the fields of forecasting, statistics, econometrics, and meteorology is by far the squared error, followed by the absolute error and the absolute percentage error. However, Hyndman and Koehler [4] argued against the general use of different metrics based on percentage errors, as they can give infinite or undefined values for certain datasets; thus, they recommended the mean absolute error (MAE) instead. Déqué [5] identified the bias, MAE, MSE, and different types of correlations as the most widespread verification metrics for deterministic forecasts of continuous variables.

The numeric values of the aforementioned accuracy measures, however, largely depend on the variability and predictability of the forecast time series data, making them unsuitable for comparing different methods verified for different datasets. A partial remedy is calculating a skill score, which reflects the percentage of error reduction achieved by a forecasting method compared to a naïve reference method. The formula of the skill score for a generic $A$ accuracy measure is

$$s = \frac{A_f - A_r}{A_p - A_r}, \tag{1}$$

where $A_f$, $A_r$, and $A_p$ stand for the accuracy of the forecast of interest, the reference forecast, and the perfect forecast, respectively.[6] With appropriate caveats, skill scores are suitable for comparing forecasts issued for different locations and time periods; therefore, the mean square error (MSE) skill score has been used widely by the meteorological forecasting community for decades.[6,7] The mean absolute scaled error (MASE), proposed by Hyndman and Koehler [4], scales the errors by the MAE of the one-step-ahead persistence forecasts, which makes it a conceptually similar but more restricted form of an MAE skill score. Driven by its benefits, the root mean square error (RMSE) skill score became the most recommended metric by Yang *et al.*[2] for solar forecasters to measure the skillfulness of forecasts. The use of the RMSE skill score is the most prevalent in solar forecasting,[8,9] but it is also popular in other forecasting disciplines, e.g., in wind power forecasting.[10] Despite its popularity, the implications of the rule of consistency on the use of skill scores have not yet been discussed in detail; therefore, this short paper revolves around the consistent calculation and interpretation of the RMSE skill score. The readers are guided through this topic using examples taken from the context of solar forecasting, but the general reasoning and recommendations also apply to other forecasting fields.

## 2 Role of consistency in forecast verification

Murphy [11] identified three types of goodness of forecasts, namely consistency, quality, and value. Consistency denotes the correspondence between the issued forecasts and the best judgments of the forecasters. Due to the inherently incomplete knowledge of the forecasters, their best judgments are uncertain and thus probabilistic by nature, which can be expressed by a predictive distribution. Insofar as deterministic forecasting is concerned, the underlying predictive distribution must be summarized into a point forecast following a directive [3]. The directive typically refers to a statistical functional, and among the various choices, the most commonly used ones are the mean and the median. In this case, consistency means that the error metric (or scoring function) used for the verification should also correspond to the directive. The mean square error (MSE) and the mean absolute error (MAE) are minimized by the mean and median of the predictive distribution, respectively. Therefore, the directive of "forecasting the mean values of the predictive distributions" is equivalent to the directive of "making MSE-optimized forecasts."

The underlying predictive densities are typically asymmetric, which means they are summarized into different point forecasts by different functionals.[12] To that end, the different directives are, except for a few special cases, conflicting: if a forecast is optimized for one directive, it will be suboptimal for other directives. Kolassa [13] argued that it makes no sense to evaluate a single set of point forecasts using different error metrics, but only by the metric that the forecasts are optimized for. To be more permissive, reporting different metrics in a forecasting paper may be useful for obtaining deeper insights into the different aspects of forecast quality, but the final verdict should always be based on the metric that aligns with the directive.

An empirical example of the conflicting nature of MSE and MAE is shown in Fig. 1, relying on the results of the physical photovoltaic (PV) power forecasting paper of Mayer and Gróf.[14] In that paper, the authors used 32,400 different physical PV model chains to create day-ahead PV power forecasts from numerical weather prediction (NWP) forecasts covering a full year with 15-min resolution. The normalized MAE and RMSE were calculated for all yearly sets of forecasts, and their scatterplot is shown in Fig. 1, where each point corresponds to a different set of forecasts. Overall, the forecasts at the top right of the figure have high errors under both metrics; therefore, they are worse than those at the bottom left. However, there is no single set of forecasts that can perform best in terms of both metrics, but there are distinct sets of MAE-optimized and MSE-optimized forecasts, along a Pareto front (denoted by red dots). The Pareto-front contains all the Pareto-efficient sets of forecasts, representing all the optimal trade-offs between the two metrics of interest. In this context, a set of forecasts is Pareto-efficient if all other sets of forecasts that are better in terms of one error metric are worse in terms of the other error metric.

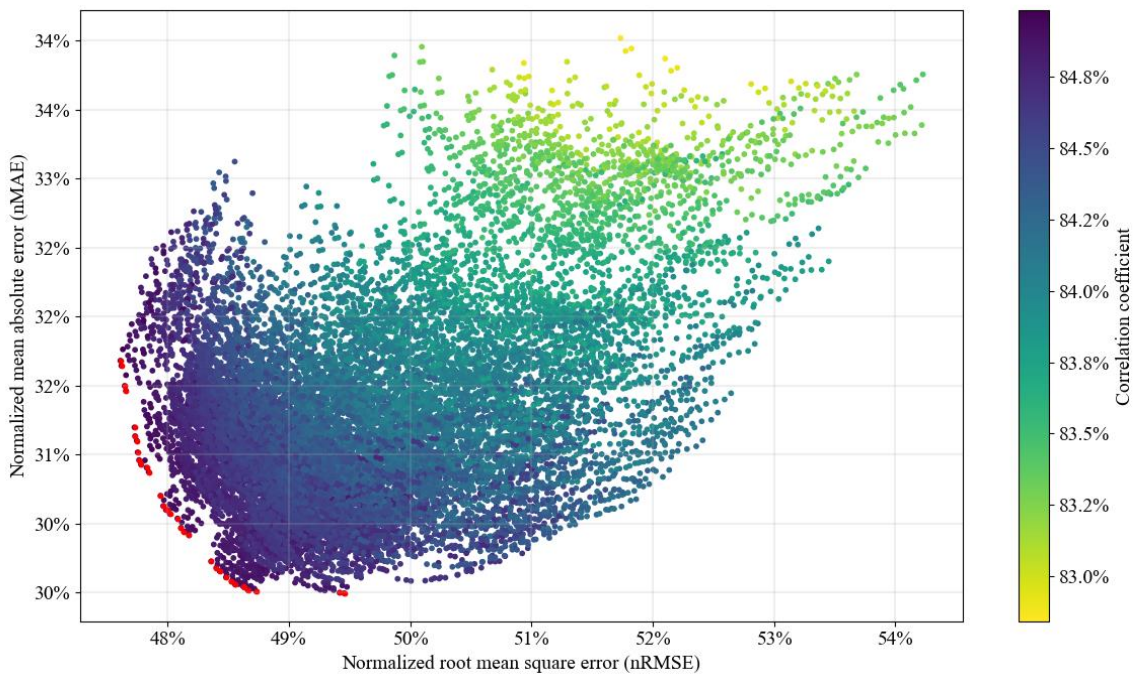

**Fig. 1.** Scatter plot of the nMAE versus the nRMSE of yearly sets of day-ahead PV power forecasts calculated by different model chains from numerical weather prediction [14]. Darker color indicates higher correlation coefficients, and the red points are the Pareto-optimal sets of forecasts.

A theoretical explanation for the conflict between the MAE and MSE has been presented by Mayer and Yang.[12] The main statistical difference between these two types of forecasts is their variance: it was proven that the MSE-optimized forecasts are always underdispersed, i.e., they lack extremely low or high forecast values, while the variance of the MAE-optimized forecasts is closer to that of the observations. Based on this, even a simple linear calibration that alters the variance of the forecast can change which metric the forecasts favor. In other words, the main statistical difference between the MSE-optimized and MAE-optimized forecasts is how they are calibrated, as reflected by their bias and variance. For this reason, the correlation coefficient between the forecasts and observations, which is not affected by bias and variance, offers a calibration-independent metric to judge the potential goodness of the forecasts.[12,15] The empirical evidence for this is shown by the color-coding of Fig. 1, where the highest correlation coefficient aligns well with the Pareto-optimal sets of forecasts, which implies that the forecasts with the best MAE and the best MSE are both awarded with a similarly high correlation coefficient.

The three linear calibration schemes proposed by Mayer and Yang [12] to create MSE-minimized, MAE-minimized, and variance-corrected forecasts, were also used by Yang et al.[16] during the verification of NWP irradiance forecasts from different sources. In this way, the raw NWP output was calibrated for different directives and then verified by the corresponding metrics, which is a good practice for getting a broad picture of the forecast quality while still respecting consistency.

In practical applications, the directive should be set to maximize the value of the forecasts. However, forecasts have no intrinsic value, but the value is only realized through their influence on the decision-making of the users.[11] Since there is a multitude of users with different goals, it is not possible to select a general "best" directive for theoretical forecasting studies where the range of the possible users is diverse or not clear. For example, four different penalty schemes are in effect for the PV power forecast errors in the different regional grids of China, which translates to four distinct directives.[17] Researchers intending to develop forecasting methods for general use are free to select the directive based on what they deem most suitable for the potential users, which largely depends on the researchers' own preference and perception of the practical needs.

In order to make a forecasting paper valuable for a wider range of practitioner audiences, it can also be useful to create and verify the forecasts, in parallel, under several directives. A good example of this is a recent paper,[18] comparing numerous different methods for hybridizing physical and machine-learning models for PV power forecasting. The author created two different sets of forecasts for all methods, one optimized for MAE while the other for MSE, and evaluated each of them with the corresponding metric. The results revealed that even the selection of method and the added value of the hybridization depends on the verification directive.

A possible way of directive-free comparisons is to create probabilistic forecasts and verify them using a strictly proper scoring rule, such as the continuous ranked probability score (CRPS).[19] A probabilistic predictive distribution can be summarized into a deterministic forecast following any directive.[3,17] However, probabilistic forecasting requires more complex and computation-intensive methods than deterministic forecasting, and many practical applications still require only deterministic forecasts. Moreover, probabilistic forecasts with lower CRPS do not always translate to deterministic forecasts with lower MSE or MAE; thus, the ranking of probabilistic forecasts can not be reliably transferred to their deterministic forecasting ability.[20]

## 3  Calibration-independent RMSE skill score calculation

Based on the above discussion, it becomes clear that the general recommendation of mandating the use of RMSE skill score in all studies contradicts the consistency rule. Calculating the RMSE skill score for forecasts issued for any directive other than minimizing the MSE leads to unfair comparisons, as the skill score is calculated from such a metric (i.e., RMSE) that the forecasts are not calibrated for. To simplify the present discussion, in what follows, the alternative directive is taken to be minimizing the MAE without loss of generality, because the MSE and MAE are the two most commonly used measures in the forecasting literature [3], although there are many other possible directives.

A seemingly good solution to resolve the inconsistency of using the RMSE skill score to evaluate MAE-optimized forecasts is to simply use the MAE skill score instead. However, as skill scores calculated from different metrics are not directly comparable,[21] this choice undermines the intended purpose of skill score, which is to provide as far as possible cross-scenario comparability. Moreover, the standard of reference for skill score calculation is the optimal convex combination of climatology and persistence (CLIPER), where the "optimal," by default, refers to the pair of convex combining weights that leads to the lowest MSE.[22–24] Calculating the MAE skill score of an MAE-optimized forecast over an MSE-optimized reference still violates consistency. Therefore, the consistent way of MAE skill score calculation is not only about changing the metrics to MAE instead of RMSE in the formula but also requires finding and using the MAE-optimized CLIPER as the reference.

An attractive alternative is to recalibrate the MAE-optimized forecasts for the lowest RMSE by a linear calibration and calculate the RMSE skill score for this set of forecasts. As shown by Mayer and Yang,[12] the RMSE of the MSE-calibrated unbiased forecasts depend only on the correlation coefficient between the forecasts and observations and the standard deviation of the observations,

$$\text{RMSE}(f, x) = \sqrt{1 - \rho^2(f, x)}\sigma(x), \qquad (2)$$

where $f$ and $x$ stand for the forecast and observation, respectively, $\rho$ is the correlation coefficient, and $\sigma$ is the standard deviation. According to Yang,[22] the RMSE of CLIPER forecasts only depends on the autocorrelation and standard deviation of the observations,

$$\text{RMSE}_{\text{cp}}(x) = \sqrt{1 - \gamma^2(h)}\sigma(x), \qquad (3)$$

where $\gamma(h)$ denotes the lag-$h$ autocorrelation of the observations.

At this juncture, one can notice the resemblance between Eqs. (2) and (3), keeping in mind that $\gamma(h)$ is the correlation coefficient between the $h$-step-ahead persistence (i.e., the lag-$h$ observations) and the observations. Based on this and the fact that climatology is the mean of the observations, CLIPER is the linearly calibrated version of the persistence for the lowest MSE.

Now, we are finally ready for the main proposal of this work. Suppose there are some forecasts $f'$ (regardless of how they are produced and what optimization strategy is used), which, after MSE-optimized linear calibration, yield $f$, then according to Eqs. (2) and (3), the RMSE skill score of $f$ depends only on the correlation coefficient and the lag-$h$ autocorrelation, i.e.,

$$s_{\text{cp}} = 1 - \frac{\text{RMSE}(f, x)}{\text{RMSE}_{\text{cp}}(x)} = 1 - \sqrt{\frac{1 - \rho^2(f', x)}{1 - \gamma^2(h)}}. \qquad (4)$$

Please note that the linear calibration does not change the correlation coefficient, i.e., $\rho(f, x) = \rho(f', x)$. One should not interpret $s_{\text{cp}}$ as the actual RMSE skill score of the raw forecasts, but to regard it as the *potential* skill score that the raw forecasts would attain if they are calibrated linearly by minimizing MSE.

The benefit of this so-called "potential RMSE skill score" is three-fold. To begin with, it is easy to calculate, as one does not need to actually perform MSE-optimized linear calibration of the forecasts. Secondly, it is directly comparable with the RMSE skill score as advised by the original formulation. Most importantly, it rewards with a high score all forecasts that have a good association with the observations, regardless of what optimization directive those forecasts initially follow.

To give perspective, a similar scatterplot as Fig. 1, but colored based on the potential RMSE skill score of Eq. (4), is shown in Fig. 2. It can be seen that the high potential RMSE skill scores not only belong to those forecasts with actually low RMSE but also to those forecasts whose RMSE would be low if they were calibrated for that metric (i.e., those points at the bottom-center of the scatter, which have low MAEs but not-so-low RMSEs). Quantitatively, the actual RMSE skill score (calculated using their actual RMSE in the numerator) of the MSE-optimized and MAE-optimized forecasts is 36.13% and 33.65%, respectively. Considering that the lowest RMSE skill score among all tested forecasts is 27.26%, this difference can be deemed significant, and as such, the actual RMSE skill score clearly suggests that the MSE-optimized forecasts are superior to MAE-optimized ones. On the other hand, however, the potential RMSE skill scores are 36.42% and 36.07% for the MSE-optimized and MAE-optimized forecasts, showing that both are almost equally good in their own way.

The highest potential RMSE skill score aligns well with the Pareto front, which means that all forecasts that represent a trade-off of the MSE-optimization and MAE-optimization directives are awarded by an almost equally high value. The potential RMSE skill score even allows ranking the forecasts within the Pareto-front; however, such ranking has little practical relevance, as from the Pareto-optimal forecasts, the choice is typically made following the directive. That said, the real significance of the potential RMSE skill score is to help identify the best forecast methods regardless of the directive, even if the different forecasts are calculated for different datasets and presented in different papers, which makes it impossible to directly plot the Pareto front.

The drawback of Eq. (4) is that it inherits the calibration-independent property from the correlation coefficient, which means that forecasts with high bias or a variance far from the observations also seem skillful. This is also visible in Fig. 2, where there are also several points with high skill scores far from the Pareto front. The forecasts represented by these points have a good association with the observations, but they are highly biased, which results in relatively high nMAE and nRMSE. For this reason, the skill score calculated using Eq. (4) must be seen as a measure of the potential rather than the actual skillfulness of the forecasts that can be achieved after a proper bias correction.

The idea of using a correlation coefficient as a skill score has already been proposed by Morf,[25] who noted that any measure of concordance could serve as a skill score. Morf recommended using Spearman's correlation coefficient directly as an alternative skill score; however, doing so makes the resulting score unsuitable for cross-scenario comparisons as it does not include any metrics related to the reference forecasts and thus undermines the main purpose of skill scores.

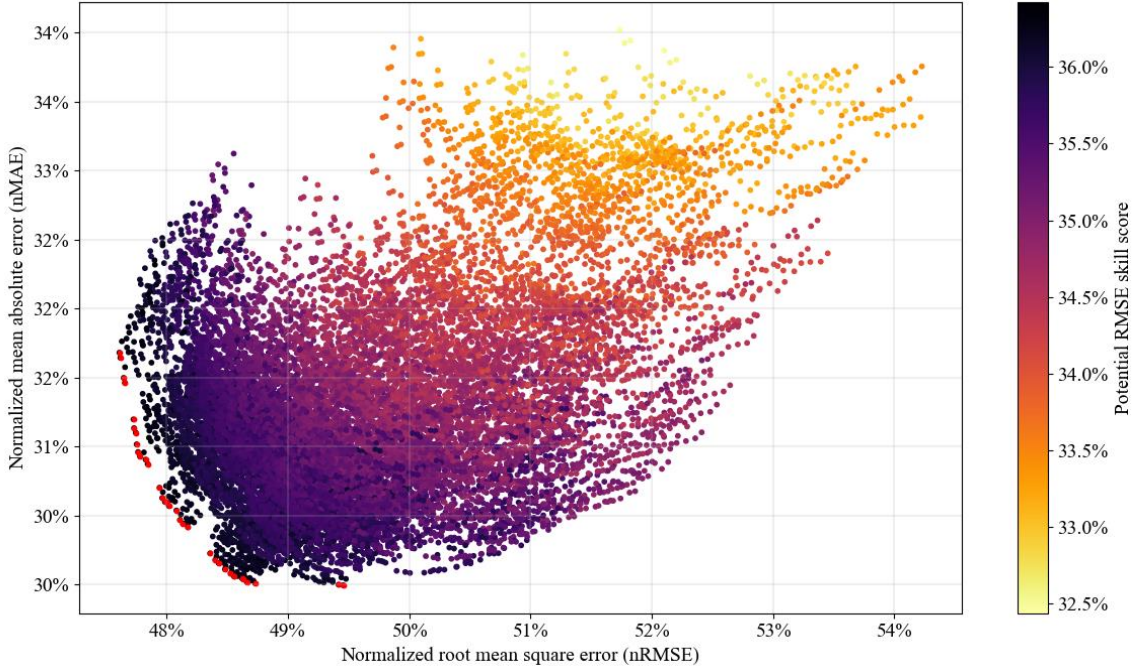

**Fig. 2.** Same as Fig. 1, but colored based on the potential RMSE skill score calculated by Eq. (4). Darker color indicates higher skill scores, and the red points are the Pareto-optimal sets of forecasts.

The discussion so far has focused on the RMSE skill score, but a potential MSE skill score can also be defined following the presented train of thought, and calculated by simply dropping the square root operator from Eq. (4), i.e.,

$$s_{\text{cp}}^{\text{MSE}} = 1 - \frac{\text{MSE}(f, x)}{\text{MSE}_{\text{cp}}(x)} = 1 - \frac{1 - \rho^2(f', x)}{1 - \gamma^2(h)}. \tag{5}$$

## 4  Conclusion

Consistency is a prerequisite for fair and meaningful forecast verification. The strict compliance to consistency, however, calls into question the use of RMSE skill score in gauging the quality of forecasts, which has hitherto been seen as one of the best metrics to compare forecasts created for different datasets. This paper provides a concise discussion on practical aspects of consistency and its possible interference with the conventional way of calculating the skill score, and more importantly, it provides a remedy for the inconsistency of using RMSE skill scores for forecasts optimized for other directives. To this end, the so-called "potential RMSE skill score" is introduced, which measures the skillfulness of forecasts if they were optimized for the mean square error. This skill score is easy to calculate, as it only relies on the crosscorrelation between the forecasts and observations and the autocorrelation of the forecasts, and it is fully comparable to the traditional formulation of the RMSE skill score. The main benefit of this new skill score over the original formula is that it much less privilege the MSE-optimized forecasts over other forecasts created under different directives. It is recommended to calculate the potential instead of the actual RMSE skill score in all papers that present deterministic forecasting methods optimized for any metric other than the MSE. The proposed new formula contributes to the fairer comparability of results and methods presented in different papers, and encourages a more thoughtful choice of directive instead of uncritically sticking with the MSE minimization. Finally, it must be noted that the potential RMSE skill score still has a slight preference toward the MSE-optimized forecasts, and it must be used with caution if the directive is significantly different from forecasting the mean or the median, e.g., it is related to a lower or higher quantile.

**Acknowledgment**

This paper was supported by the National Research, Development and Innovation Fund, project no. OTKA-FK 142702 and by the Hungarian Academy of Sciences through the Sustainable Development and Technologies National Programme (FFT NP FTA) and the János Bolyai Research Scholarship.

**Author Declarations**

*Conflict of Interest*

The authors have no conflicts to disclose.

*Author Contributions*

**Martin János Mayer**: Conceptualization, Methodology, Investigation, Data curation, Visualization, Writing – original draft. **Dazhi Yang**: Methodology, Validation, Writing – review & editing.

**Data availability**

Data sharing is not applicable to this article as no new data were created or analyzed in this study.